\documentclass[twocolumn]{aastex63}

\usepackage{xcolor}
\definecolor{green2}{rgb}{0.0,0.5,0}

\submitjournal{AJ}

\shorttitle{Spiral in the disk of CQ Tau}
\shortauthors{Safonov et al.}
\graphicspath{{./}{figures/}}

\usepackage{amsmath}

\begin{document}

\title{Apparent motion of the circumstellar envelope of CQ Tau in scattered light}

\correspondingauthor{Boris Safonov}
\email{safonov@sai.msu.ru}

\author[0000-0003-1713-3208]{Boris S. Safonov}
\affiliation{Sternberg Astronomical Institute Lomonosov Moscow State University \\ 119992, Universitetskij prospekt 13, Moscow, Russian Federation}

\author[0000-0003-0647-6133]{Ivan A. Strakhov}
\affiliation{Sternberg Astronomical Institute Lomonosov Moscow State University \\ 119992, Universitetskij prospekt 13, Moscow, Russian Federation}

\author[0000-0003-2228-7914]{Maria V. Goliguzova}
\affiliation{Sternberg Astronomical Institute Lomonosov Moscow State University \\ 119992, Universitetskij prospekt 13, Moscow, Russian Federation}

\author[0000-0001-7177-0609]{Olga V. Voziakova}
\affiliation{Sternberg Astronomical Institute Lomonosov Moscow State University \\ 119992, Universitetskij prospekt 13, Moscow, Russian Federation}



\begin{abstract}

The study of spiral structures in protoplanetary disks is of great importance for understanding of processes in the disks, including planet formation. Bright spiral arms were detected in the disk of young star CQ~Tau by \citet{Uyama2020} in $H$ and $L$ bands. The spiral arms are located inside the gap in millimeter size dust, recovered earlier using ALMA observations \citep{UbeiraGabellini2019}. To explain the gap, \citet{UbeiraGabellini2019} proposed the existence of a planet with the semimajor axis of 20~AU.

We obtained multiepoch observations of a spiral feature in the circumstellar envelope of CQ~Tau in $I_c$ band using a novel technique of differential speckle polarimetry. The observations covering period from $2015$ to $2021$ allow us to estimate the pattern speed of spiral: $-0.2\pm1.1^{\circ}$/yr (68\% credible interval, positive value indicates counter--clockwise rotation), assuming face--on orientation of the disk. This speed is significantly smaller than expected for a companion--induced spiral, if the perturbing body has the semimajor axis of 20~AU. We emphasize that the morphology of the spiral structure is likely to be strongly affected by shadows of a misaligned inner disk detected by \citet{Eisner2004}.
\end{abstract}



\section{Introduction}
\label{sec:intro}

Scattered light observations of some protoplanetary disks at high angular resolution revealed spiral structures with characteristic sizes comparable to or several times larger than the Solar system size \citep{Grady1999,Muto2012,Grady2013,Benisty2015,Muro-Arena2020}. Models predict the spiral waves to emerge due to the influence of a companion \citep{Juhasz2015,Dong2016} or due to a gravitational instability of the disk \citep{Dong2015}. Perhaps the best tool to discriminate between those hypotheses is to measure the pattern speed of the spiral structure using observations at multiple epochs \citep{Ren2020}. In the case of a companion--induced spiral the pattern speed equals to the Keplerian speed of the companion. At the same time, a gravitational instability spiral wave moves at the local Keplerian speed at all distances from the star. 

Given the origin of a spiral is known, the analysis of its observational parameters can be indispensable for constraining the conditions in the disk. It can help to determine the position and mass of the companion, which is still embedded in the disk and therefore may be difficult to detect by direct techniques \citep{Zhu2015,Fung2015}. This is especially relevant as long as the observations of exoplanets in the process of formation are still rare (see PDS70b for exception \citep{Keppler2018}). Gravitational instability induced waves can be used to estimate the mass of the disk, one of its most important parameters \citep{Dong2015}.

The interpretation of the spiral structures observations has to take into account the effects of uneven and variable illumination, caused by disk self--shadowing \citep{Ren2020,Xie2021}. These include global variations of brightness of the whole arms (e.g. MWC758 \citep{Benisty2015}) or narrow shadow lanes stretching in the radial direction (e.g.  HD135344B, \citet{Stolker2016}), or both (e.g. HD100453, \citet{Benisty2017,Long2017}). Moreover, the shadows are expected to create pressure gradients strong enough to induce the spiral density waves, as was demonstrated by \citet{Montesinos2016}. In this case their pattern speed is expected to be the same as in the case of spirals induced by the gravitational instability \citep{Montesinos2018}.



Bright spiral features with characteristic projected size of $25-65$~AU have been recently detected in the circumstellar disk of YSO CQ~Tau ($\alpha=05^h35^m58^s.47$, $\delta=+24^{\circ}44^{\prime}54^{\prime\prime}.1$, $d=149.4\pm1.3$~pc, \citep{GaiaEDR3}) in scattered polarized light ($H$--band) and in total intensity ($L^{\prime}$--band) by \citet{Uyama2020}. CQ~Tau is an intermediate mass HAeBe star (Spectral type: F2, age: 10~Myr, $M=1.67M_\odot$, \citet{UbeiraGabellini2019}) and a well--known UX~Ori variable \citep{Berdyugin1990} located in the Taurus star forming region. Here we adopt distance of $d=163\pm2$~pc from \citep{GAIADR2} for the consistent comparison with previous works.


A prominent depression was found in the gas and mm--sized dust  distribution of the CQ Tau circumstellar disk by \citet{UbeiraGabellini2019} using high angular resolution ALMA observations in 1.3 mm continuum and CO lines. According to \citet{UbeiraGabellini2019}, the inclination and the position angle of the major axis of the gas disk are $35^{\circ}$ and $55^{\circ}$, respectively. The sizes of gas and large dust (1 $\mu$m$-1$ cm) cavities are $\approx20$~AU and $\approx52$~AU, respectively. \citet{UbeiraGabellini2019} proposed that the cavity is cleared by a planet with a mass of $6-9M_\mathrm{Jup}$ and a semi--major axis of 20~AU. Taking into account the mass of the star $1.67M_\odot$ the expected orbital period is 75~yr.

The existence of a planet is further supported by the spiral departures from Keplerian kinematics of the gas found by \citet{Wolfer2020}. \citet{Uyama2020} reported an excess brightness of the spiral in $L^{\prime}$--band and proposed that it may be explained by an additional thermal emission due to a disk--planet interaction. The possible connection between the spiral and the putative planet can be tested by measurement of spiral pattern speed, such analysis is favoured for CQ Tau by the short expected orbital period of the planet.

The SED of CQ~Tau shows a significant NIR excess, which requires the presence of small dust particles in the cavity \citep{UbeiraGabellini2019}. The inner disk with typical size of 0.45~AU around the star was detected in $K$--band using Palomar Testbed Interferometer \citep{Eisner2004}. These facts and UX~Ori variability indicate that the inner part of the circumstellar disk may affect the illumination conditions of the outer disk, which should be taken into account during analysis.


\begin{figure*}[t]
\centering
\includegraphics[width=16cm]{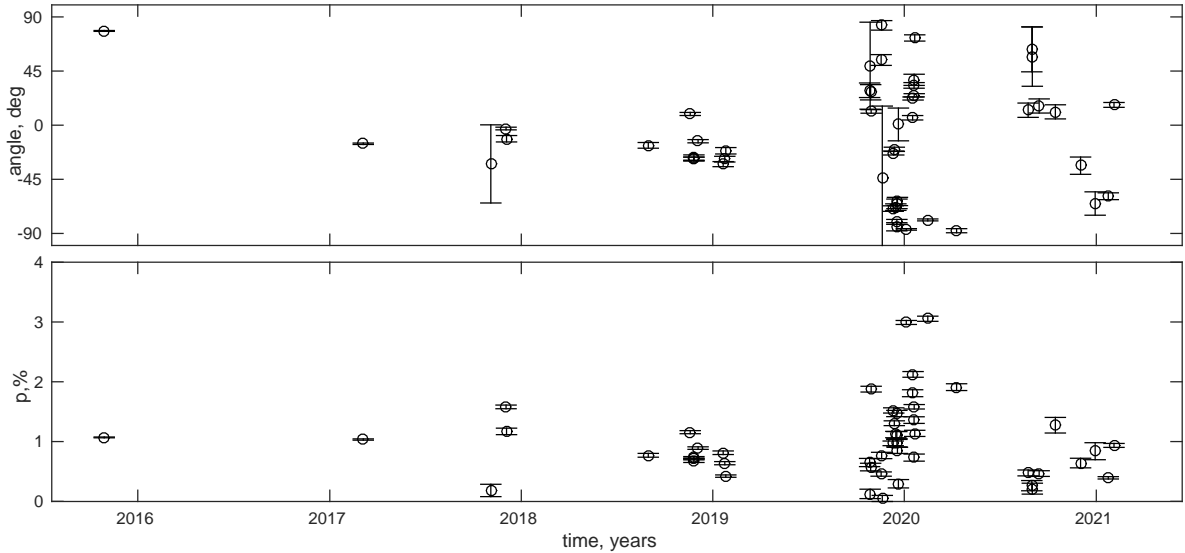}
\caption{Top and bottom panels contain the dependence of polarization angle and fraction of CQ~Tau on time, respectively.
\label{fig:groups}}
\end{figure*}

In the current paper, we present observations of the circumstellar envelope of CQ~Tau using differential speckle polarimetry (DSP) at 48 epochs distributed over the period of more than 5 years in $I_c$--band (effective wavelength is 0.806~$\mu$m). DSP is a novel technique based on analysis of series of short--exposure frames obtained through a dual--beam polarimeter without an adaptive optics correction. The spiral structure with the characteristic size of $0.1-0.15^{\prime\prime}$, which corresponds to $15-25$~AU at the distance of the source, was detected in the majority of cases; our observations allow us to trace the evolution of its morphology. The work is organized as follows. Sections \ref{sec:observations} and \ref{sec:processing} contain the description of the observational data and processing methods, respectively. In the section \ref{sec:results}, we compare our data with those available in the literature and investigate the variability of the morphology of the spiral structure of CQ~Tau. The conclusions are given in section \ref{sec:conclusions}.

\section{Observations}
\label{sec:observations}

The observations were made using SPeckle Polarimeter (SPP), which is a dual--beam polarimeter and speckle camera equipped with an Electron Multiplying CCD (EMCCD) sensor Andor iXon 897. The instrument is installed at the $2.5$-m telescope of Caucasian Mountain Observatory of Sternberg Astronomical Institute of Lomonosov Moscow State University \citep{Kornilov2014}. Splitting of the orthogonally polarized beams is done by a Wollaston prism, both images formed by the latter are located side--by--side on the same detector, see Fig.~\ref{fig:DSPfirstSteps}. The instrument includes a rotating half--wave plate (HWP) for the modulation of the polarization state of incoming radiation. The atmospheric dispersion is compensated by a dedicated unit consisting of two rotatable direct vision Amici prisms. The plate scale of the instrument is $0.0206^{\prime\prime}$/pix. The plate scale determination and calibration of the position angle of the instrument is performed by observations of wide binaries, as described by \citet{Safonov2017}. The field of view of the instrument is a rectangle with the dimensions of $5^{\prime\prime}\times10^{\prime\prime}$.

The SPP can operate in several bands of visible wavelength range, although in the present work we use the observations in the $I_\mathrm{c}$ band only, at an effective wavelength $\lambda=806$~nm. At the wavelengths longer than $460$~nm, including the whole $I_\mathrm{c}$ band, the detector oversamples the diffraction limited point spread function of the feeding telescope, which has an aperture diameter $D=2.5$~m. The main detector of SPP has a negligible effective readout noise, smaller than 0.1 e$^{-}$, and can run at high frame rates very close to 100\% duty cycle thanks to the frame transfer technology. These features of the SPP, and other EMCCD--based speckle interferometers in general, ensure the reliable detection of speckle structure of the images degraded by the atmospheric optical turbulence, which subsequently allows the computation of the Fourier spectra of these images. In our case the resulting Fourier transforms are used as an input for the differential speckle polarimetry (DSP) algorithm. The detailed description of the instrument design and calibration procedures is presented in paper \citep{Safonov2017}, which also discusses the implementation of the speckle interferometry and integral polarimetry with the SPP.

The observations for the present study were obtained in so called fast polarimetry mode. In this mode the detector obtains a series of short--exposure images at 33 frames per second, which gives exposure time of 30~ms for a single frame. At the same time the HWP continuously rotates at $300^{\circ}$/sec. The rotation of the HWP serves two purposes. Firstly, it allows to measure both Stokes parameters corresponding to linear polarization. Secondly, it switches the images corresponding to orthogonal polarizations in the beams of the polarimeter each 5 frames, allowing for application of double difference technique \citep{Bagnulo2009}. The double difference discards most of the instrumental effects associated with the differences between two channels of the polarimeter.

In total we obtained 48 series of CQ Tau in the period from October 29th, 2015 to February, 2nd, 2021, all of them in $I_c$ band, they are listed in Table~\ref{tab:obslog}. The series duration varied from 2000 to 60000 frames (11000-12000 frames in most cases). The integral total flux polarimetry extracted from these observations is presented in Fig.~\ref{fig:groups} and Table~\ref{tab:obslog}. Some of observations were secured while the instrument was installed in the Cassegrain focus, and some in the Nasmyth focus, as indicated in the observational log, Table~\ref{tab:obslog}. For the observations in the Nasmyth focus the polarimetric measurements were corrected for the instrumental polarization by the application of the inverse Mueller matrix of the telescope, see details in \citep{Safonov2017}. In some cases we estimated the magnitude by quasi--simultaneous observations of the standard star HIP25001.

\section{Processing}
\label{sec:processing}

\subsection{Differential polarimetric visibility estimation}

\begin{figure}[t]
\centering
\includegraphics[width=8cm]{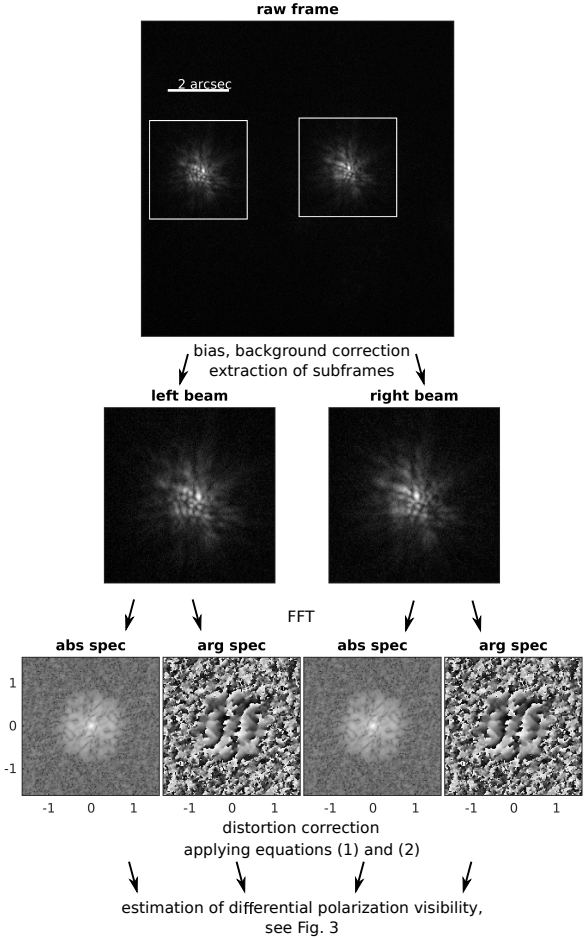}
\caption{Scheme illustrating the first steps of the differential speckle polarimetry processing using a single frame as an example. The first row displays the raw frame with two images corresponding to orthogonal polarizations. In the second row the extracted subframes are presented. Third row shows the Fourier transforms of the subframes. First and third panel contain the amplitude, second and fourth contain the argument of the complex Fourier transform.}
\label{fig:DSPfirstSteps}
\end{figure}

All observations were processed by the DSP method. For each frame processing started with the subtraction of bias and background. Then the images $F_L$, $F_R$ corresponding to the beams of the polarimeter were extracted and their Fourier transforms $\widetilde{F}_{L,i}(\boldsymbol{f})$ and $\widetilde{F}_{R,i}(\boldsymbol{f})$ were computed. Here $L$ and $R$ indices mean left and right beam of the polarimeter. $i$ is the frame number in series, $\boldsymbol{f}$ is the vector of spatial frequency. The distortion of Wollaston prism and atmospheric dispersion compensator prisms was corrected in the Fourier space. These steps are illustrated in Fig.~\ref{fig:DSPfirstSteps}

The resulting spectra were combined in the following way (we omit the dependence on spatial frequency $\boldsymbol{f}$ for brevity):
\begin{equation}
\mathcal{R}_{ch} = 1+\frac{\bigl\langle (\widetilde{F}_{L,i} - \widetilde{F}_{R,i})  ( \widetilde{F}_{L,i} + \widetilde{F}_{R,i})^*\cos(h \theta_i) \bigr\rangle_i}{\bigl\langle (\widetilde{F}_{L,i} + \widetilde{F}_{R,i})  (\widetilde{F}_{L,i} + \widetilde{F}_{R,i})^* \bigr\rangle_i - N_e^{-1}},
\label{eq:RavgC}
\end{equation}
\begin{equation}
\mathcal{R}_{sh} = 1+\frac{\bigl\langle (\widetilde{F}_{L,i} - \widetilde{F}_{R,i})  (\widetilde{F}_{L,i} + \widetilde{F}_{R,i})^*\sin(h \theta_i) \bigr\rangle_i}{\bigl\langle ( \widetilde{F}_{L,i} + \widetilde{F}_{R,i})  (\widetilde{F}_{L,i} + \widetilde{F}_{R,i})^* \bigr\rangle_i - N_e^{-1}}.
\label{eq:RavgS}
\end{equation}
Here $\theta_i$ is the position angle of the HWP for the frame $i$. $^{*}$ stands for the complex conjugation. The parentheses $\langle*\rangle$ indicate averaging  over frames in series. These equations represent the demodulation of signal, initially modulated by rotation of the HWP. They decompose the signal into harmonics $h$. As long as the detector obtains $\approx40$ frames per revolution of the HWP, we calculated values (\ref{eq:RavgC}) and (\ref{eq:RavgS}) up to $N_h=20$. $N_{e}$ is the average number of photons in a single frame.

The general idea of equations (\ref{eq:RavgC},\ref{eq:RavgS}) is to average the cross--spectra of images corresponding to Stokes $Q$ and $I$ and subsequently normalize them by the average power spectrum of Stokes $I$ image. Previously in \citep{Safonov2019} (hereinafter S19) we demonstrated that $R_{ch}$ and $R_{sh}$ corresponding to $h=4$ can be considered as an estimators of differential polarimetric visibility (DPV) $\mathcal{R}_Q$ and $\mathcal{R}_U$:
\begin{equation}
\mathcal{R}_{c4}(\boldsymbol{f}) = \mathcal{R}_{Q,\mathrm{ins}}(\boldsymbol{f})\mathcal{R}_Q(\boldsymbol{f}),
\end{equation}
\begin{equation}
\mathcal{R}_{s4}(\boldsymbol{f}) = \mathcal{R}_{U,\mathrm{ins}}(\boldsymbol{f})\mathcal{R}_U(\boldsymbol{f}).
\end{equation}
$R_{ch}$ and $R_{sh}$ for $h=0$ characterize the optics after the HWP. Other harmonics ($h\ne0$ and $h\ne4$) are expected to be zero and thus can be used for the noise estimation (S19, appendix B). The factors $\mathcal{R}_{Q,\mathrm{ins}}(\boldsymbol{f})$ and $\mathcal{R}_{U,\mathrm{ins}}(\boldsymbol{f})$ characterize the effect of instrumental polarization, they are discussed in below. 

The DPV is defined by equations:
\begin{equation}
\mathcal{R}_Q(\boldsymbol{f}) = \frac{\widetilde{I}(\boldsymbol{f})+\widetilde{Q}(\boldsymbol{f})}{\widetilde{I}(\boldsymbol{f})-\widetilde{Q}(\boldsymbol{f})},\,\,\,\mathcal{R}_U(\boldsymbol{f}) = \frac{\widetilde{I}(\boldsymbol{f})+\widetilde{U}(\boldsymbol{f})}{\widetilde{I}(\boldsymbol{f})-\widetilde{U}(\boldsymbol{f})}.
\label{eq:Rdef}
\end{equation}
Here $\widetilde{I}, \widetilde{Q}, \widetilde{U}$ are the Fourier transforms of the Stokes parameters distributions in the object. One can see that DPV is the ratio of visibilities of the object in orthogonal polarizations and can be defined for two Stokes parameters describing linear polarizations $Q$ and $U$.

\begin{figure*}[t]
\centering
\includegraphics[width=16.5cm]{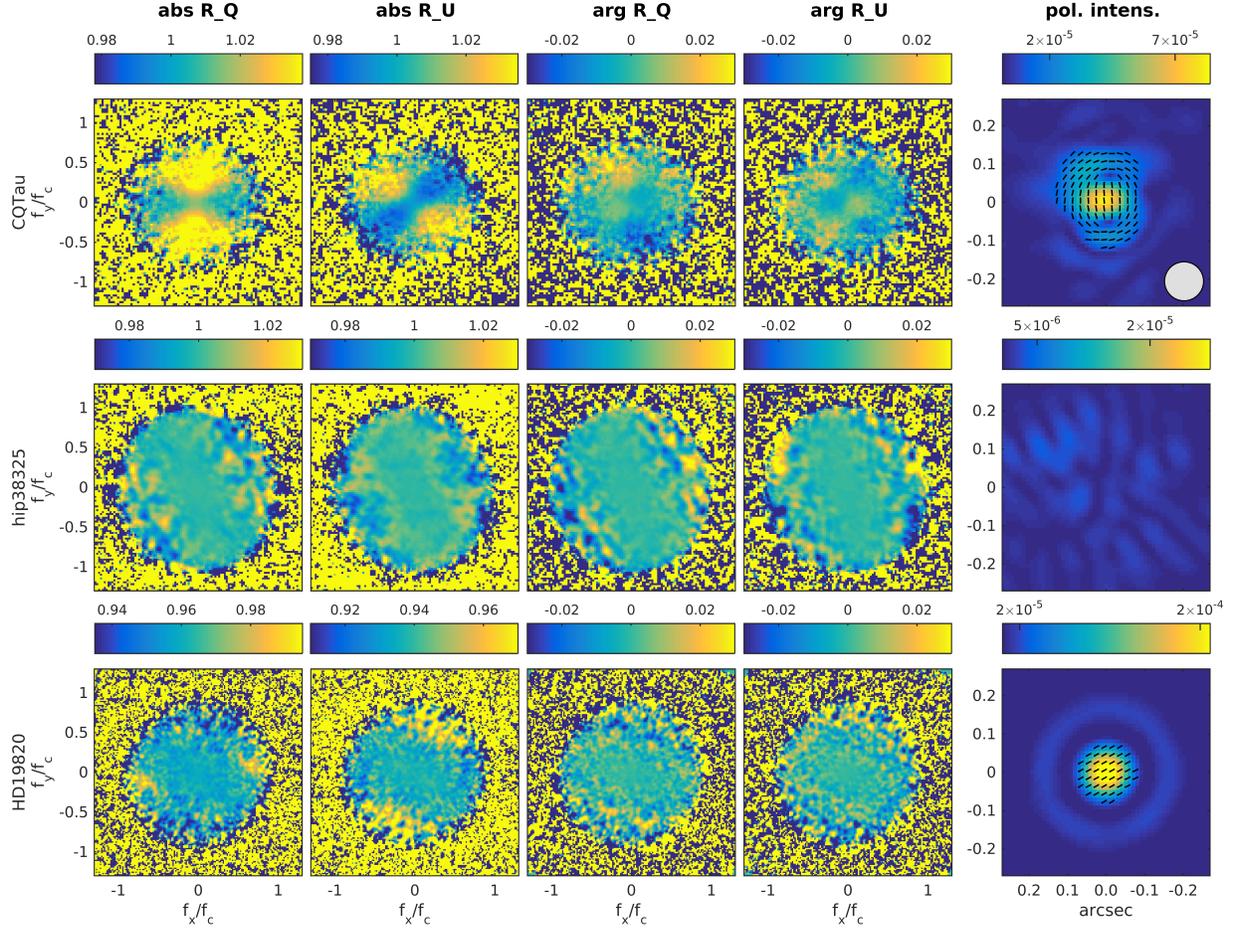}
\caption{Columns 1-4: the measurements of $\mathcal{R}_Q$ and $\mathcal{R}_U$ for CQ Tau, unpolarized star HIP38325, polarization standard HD19820, from the top to the bottom. Columns 1-2 contain the absolute part of signal, the columns 3-4 contain its argument. The axes are shown in spatial frequency, normalized by cut-off frequency $\lambda/D$, where $D=2.5$~m is the aperture diameter. The rightmost column presents the image of the envelope in the polarized intensity relative to the total flux of the star, reconstructed by the algorithm presented in section~\ref{sec:polimg}. Dashes indicate the orientation of polarization. Please note different brightness scales, the unit is polarized flux per pixel divided by the total non--polarized flux from the star. Grey circle in the upper right image indicates typical angular resolution. In all images the north is up, the east is left.}
\label{fig:rvals}
\end{figure*}

The measurement of modulus of $\mathcal{R}_Q$ and $\mathcal{R}_U$ was demonstrated by \citet{Norris2012} using adaptive optics assisted aperture masking at NaCo/VLT. Unlike that experiment, in SPP the full pupil of the 2.5-m telescope is employed, which provides larger throughput of the system. Apart from this, no adaptive optics correction is used in our case. The signal in the Fourier spectrum of the image at high frequencies is lower than it would be with such a correction. On the other hand, the optical system is simpler and more suitable for the precision polarimetry. Note that DSP allows one to obtain not only modulus, but also the argument of $\mathcal{R}_Q$ and $\mathcal{R}_U$.

The quantities in equations (\ref{eq:RavgC}-\ref{eq:Rdef}) are defined in the instrument reference system. The formulae for their conversion in horizontal and equatorial reference frames are given in S19, appendix B. The correction for instrumental terms $\mathcal{R}_{Q,\mathrm{ins}}$ and $\mathcal{R}_{U,\mathrm{ins}}$ is done in horizontal reference frame, as long as these factors are associated with the telescope, which has an alt-az mount. This operation is described in Appendix \ref{app:instrpol}. Then $\mathcal{R}_Q$ and $\mathcal{R}_U$ are converted to equatorial reference frame. For observations secured in the Cassegrain focus the instrumental terms $\mathcal{R}_{Q,\mathrm{ins}}$ and $\mathcal{R}_{U,\mathrm{ins}}$ are closer than $10^{-4}$ to unity, much smaller than the typical level of signal for CQ Tau. Therefore the correction for the instrumental polarization effect was not applied for this focus.

As an example of measurement of $\mathcal{R}_Q$ and $\mathcal{R}_U$, we provide the data for CQ~Tau obtained in the Cassegrain focus at UT 2015-10-29 00:51 in Fig.~\ref{fig:rvals}.  Here the DPV can be traced to frequencies $0.7-0.8f_c$, where $f_c$ is the cut--off frequency $D/\lambda$. $D=2.5$~m is the telescope aperture diameter, $\lambda=0.806$~$\mu$m is the wavelength of the observation. Hence the cut--off frequency $f_c=3.1\times10^{6}$~RAD$^{-1}$ in our case. 

In order to characterize the quality of data we estimated the noise of $\mathcal{R}_Q$ and $\mathcal{R}_U$ using method presented in S19. Then we average the noise in an annular region of Fourier space limited by $0.3f_c<|\boldsymbol{f}|<0.4f_c$. For the series obtained at 2015-10-29 00:51 for CQ Tau it is $6.5\times10^{-3}$ RAD, in terms of RMS. This quantity computed for all series is presented in seventh column of Table~\ref{tab:obslog}.

The Fig.~\ref{fig:rvals} also contains measurement for the unpolarized star HIP38325 secured at UT 2015-10-30 02:40 and the polarization standard star HD19820 secured at UT 2019-12-12 20:16. In both cases the $I_c$ band was used. Both stars have no circumstellar dusty envelopes.

One can see that $\mathcal{R}_Q$ and $\mathcal{R}_U$ for CQ~Tau significantly deviate from the constant level. This behaviour is  different from the unpolarized star and polarization standard star. The structure of $|\mathcal{R}_Q|$ and $|\mathcal{R}_U|$ for CQ~Tau exhibits butterfly pattern, which indicates scattering envelope, as in the case of W~Hya, reported by \citet{Norris2012}. Interestingly, there is significant structure in the argument of $\mathcal{R}_Q$ and $\mathcal{R}_U$ as well, favouring an asymmetry of the polarized flux distribution.

\subsection{Reconstruction of the image of the envelope in polarized intensity}
\label{sec:polimg}

If the object can be represented as a sum of a point-like star and an extended envelope, the Fourier transforms of Stokes parameters distributions will have the following form:
\begin{equation}
\widetilde{I}(\boldsymbol{f}) = \widetilde{I}_\star + \widetilde{I}_\mathrm{e}(\boldsymbol{f}),
\label{eq:objectI}
\end{equation}
\begin{equation}
\widetilde{Q}(\boldsymbol{f}) = q_\star\widetilde{I}_\star + \widetilde{Q}_\mathrm{e}(\boldsymbol{f}),
\label{eq:objectQ}
\end{equation}
\begin{equation}
\widetilde{U}(\boldsymbol{f}) = u_\star\widetilde{I}_\star + \widetilde{U}_\mathrm{e}(\boldsymbol{f}).
\label{eq:objectU}
\end{equation}
Here $\widetilde{I}_\star$ is the Fourier transform of the image of the star, a value equal to the flux from the star $I_\star$ for all frequencies. $q_\star$, $u_\star$ are the dimensionless Stokes parameter of the star. $\widetilde{I}_\mathrm{e}(\boldsymbol{f})$, $\widetilde{Q}_\mathrm{e}(\boldsymbol{f})$, $\widetilde{U}_\mathrm{e}(\boldsymbol{f})$ --- Fourier transforms of the envelope image in Stokes parameters $I$, $Q$, $U$, respectively.

In the following we take into account that the polarization of the star is small $q_\star\ll1$, $u_\star\ll1$, and the envelope is faint in comparison with the star $\widetilde{Q}_\mathrm{e}\ll\widetilde{I}_\star$, $\widetilde{U}_\mathrm{e}\ll\widetilde{I}_\star$. Substituting (\ref{eq:objectI}-\ref{eq:objectU}) into (\ref{eq:Rdef}) and keeping only first order terms, we obtain for the Fourier transforms of the envelope flux distributions:
\begin{equation}
\widetilde{Q}_\mathrm{e}^{\prime}(\boldsymbol{f}) = 0.5\bigl(\mathcal{R}_Q(\boldsymbol{f})-1\bigr),
\end{equation}
\begin{equation}
\widetilde{U}_\mathrm{e}^{\prime}(\boldsymbol{f}) = 0.5\bigl(\mathcal{R}_U(\boldsymbol{f})-1\bigr).
\end{equation}
Here the quantities on the left side of equations are normalized by the total stellar flux and additionally include the contribution of polarized direct stellar radiation:  $\widetilde{Q}_\mathrm{e}^{\prime}(\boldsymbol{f})=(\widetilde{Q}_\mathrm{e}(\boldsymbol{f})/I_\star)+q_\star$, $\widetilde{U}_\mathrm{e}^{\prime}(\boldsymbol{f})=(\widetilde{U}_\mathrm{e}(\boldsymbol{f})/I_\star)+u_\star$.

As one can see from Fig.~\ref{fig:rvals}, $\mathcal{R}_Q$ and $\mathcal{R}_U$, and subsequently $\widetilde{Q}_\mathrm{e}^{\prime}(\boldsymbol{f})$ and $\widetilde{U}_\mathrm{e}^{\prime}(\boldsymbol{f})$ have acceptable signal--to--noise ratio up to frequencies 0.7--0.8$f_c$. Therefore we multiplied them by low--pass filter in order to suppress the wings of point spread function in the image space and reduce the effect of the noise. For this spatial filter we used the optical transfer function (OTF) of a circular aperture:
\begin{equation}
G(\boldsymbol{f})=(2/\pi)\Bigl[\mathrm{arccos}z-z\sqrt{1-z^2} \Bigr],
\end{equation}
where $z=2|\boldsymbol{f}|\lambda/D^{\prime}$. We took $D^{\prime}=0.7D$, where D=2.5~m, for all the data. 

$\widetilde{Q}_\mathrm{e}^{\prime}(\boldsymbol{f})G(\boldsymbol{f})$ and $\widetilde{U}_\mathrm{e}^{\prime}(\boldsymbol{f})G(\boldsymbol{f})$ are defined in the region $|f_x|<f_s/2$, $|f_y|<f_s/2$ in the Fourier space. Here $f_s=1/\alpha_s$ is the sampling frequency, $\alpha_s$ is the angular scale of the detector. Therefore the image produced by applying inverse Fourier transform to them will have the pixel size of $\alpha_s$. We artificially increased the region in which $\widetilde{Q}_\mathrm{e}^{\prime}(\boldsymbol{f})G(\boldsymbol{f})$ and $\widetilde{U}_\mathrm{e}^{\prime}(\boldsymbol{f})G(\boldsymbol{f})$ are defined two times by padding it with zeros. This operation did not add or remove any information from the data, however the pixel size in resulting image decreased to $\alpha_s/2$, making the appearance of the object in the image space smoother.

Filtered and padded $\widetilde{Q}_\mathrm{e}^{\prime}(\boldsymbol{f})$ and $\widetilde{U}_\mathrm{e}^{\prime}(\boldsymbol{f})$ were transformed into the image space by applying inverse FFT. This reconstruction process results in the estimation of polarized intensity in the circumstellar environment $Q^{\prime}(\boldsymbol{\alpha})$ and $U^{\prime}(\boldsymbol{\alpha})$, where $\boldsymbol{\alpha}$ is a two--dimensional vector of spatial coordinate relative to the point--like unpolarized star. The quantities $Q^{\prime}(\boldsymbol{\alpha})$ and $U^{\prime}(\boldsymbol{\alpha})$ are then used for the computation of the polarized intensity $I_p(\boldsymbol{\alpha})$ and the angle of polarization $\chi(\boldsymbol{\alpha})$:
\begin{equation}
I_p(\boldsymbol{\alpha}) = \sqrt{Q^{\prime2}(\boldsymbol{\alpha})+U^{\prime2}(\boldsymbol{\alpha})},
\end{equation}
\begin{equation}
\chi(\boldsymbol{\alpha}) = \frac{1}{2}\mathrm{arctg}\bigl(U^{\prime}(\boldsymbol{\alpha})/Q^{\prime}(\boldsymbol{\alpha})\bigr).
\end{equation}
Recall that the polarized intensity is the product of the total intensity and the fraction of polarization.

The examples of $I_p(\boldsymbol{\alpha})$ and $\chi(\boldsymbol{\alpha})$ are presented in Fig.~\ref{fig:rvals} in the rightmost column. As long as spatial filter is non--zero up to frequencies $0.7f_c\equiv2.17\times10^6$ RAD$^{-1}$, the reconstructed image of the circumstellar envelope has typical resolution of $\approx0.1^{\prime\prime}$. For the unpolarized star HIP38325 the image does not contain any significant structure. On the other hand,  in the image of CQ Tau envelope, a spiral arm in $0.12^{\prime\prime}$ to the North from the star is present. There is also a less prominent feature in $\lesssim0.1^{\prime\prime}$ to the South from the star. Both features demonstrate an azimuthal polarization pattern which indicates that they are parts of a reflection nebula associated with the star.

Note that the estimations $Q^{\prime}(\boldsymbol{\alpha})$ and $U^{\prime}(\boldsymbol{\alpha})$ contain the contribution from the polarized direct radiation of the star. This effect is prominent in the case of polarization standard HD19820. In the bottom right panel of Fig.~\ref{fig:rvals} one can see a bright source of polarized radiation coinciding with the star. In case of CQ~Tau there is such a source as well, which is discussed in section~\ref{sec:temporal}.

\section{Results}
\label{sec:results}

\subsection{Comparison with data from literature}

The fact that the northern spiral is brighter and farther from the star than the southern one suggests that it lies on the far edge of the flared disk, which has more favorable conditions for scattering the stellar radiation. This assumption is supported by the kinematics data of \citet{Wolfer2020}, which show that the disk rotates counter--clockwise and the northwest edge is farther.

\begin{figure}[t]
\centering
\includegraphics[width=8cm]{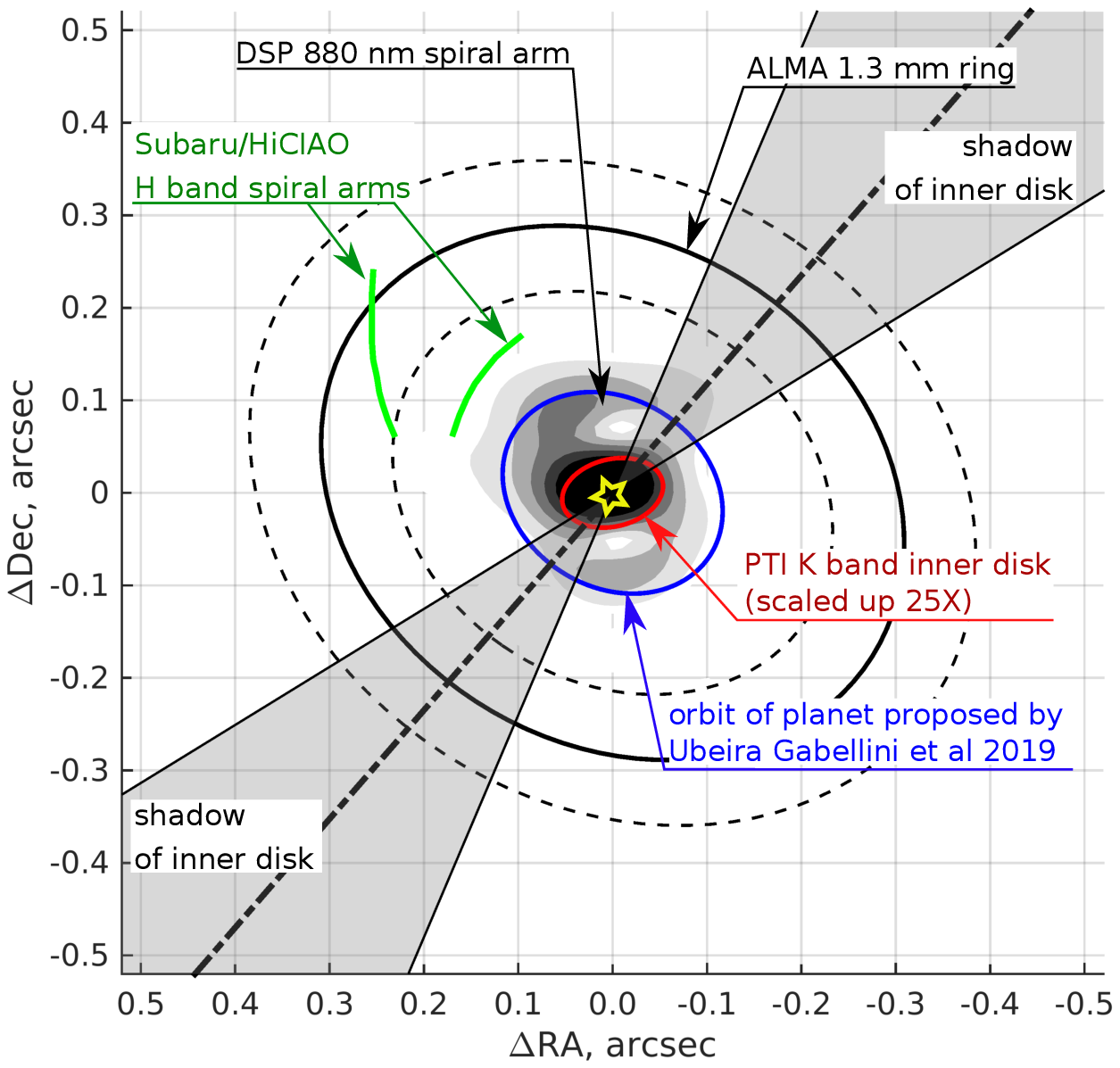}
\caption{The comparison of our data (indicated by grey contours) with the results from the literature. Green line is the spiral arm found using Subaru/HiCIAO \citep{Uyama2020}. Black ellipse is the 1.3 mm continuum ring obtained using ALMA by \citet{UbeiraGabellini2019} (dashed line indicate the approximate borders of this ring). Blue line indicates a possible orbit of the planet proposed by \citet{UbeiraGabellini2019}. Red ellipse shows the orientation and inclination of the inner disk according to PTI data \citep{Eisner2004} (enlarged 25 times for a visual clarity). Grey segments is the assumed shadow cast by the inner disk on the outer ring, see section \ref{sec:shadow}. 
\label{fig:shadow}}
\end{figure}

\begin{figure*}[t]
\centering
\includegraphics[width=18cm]{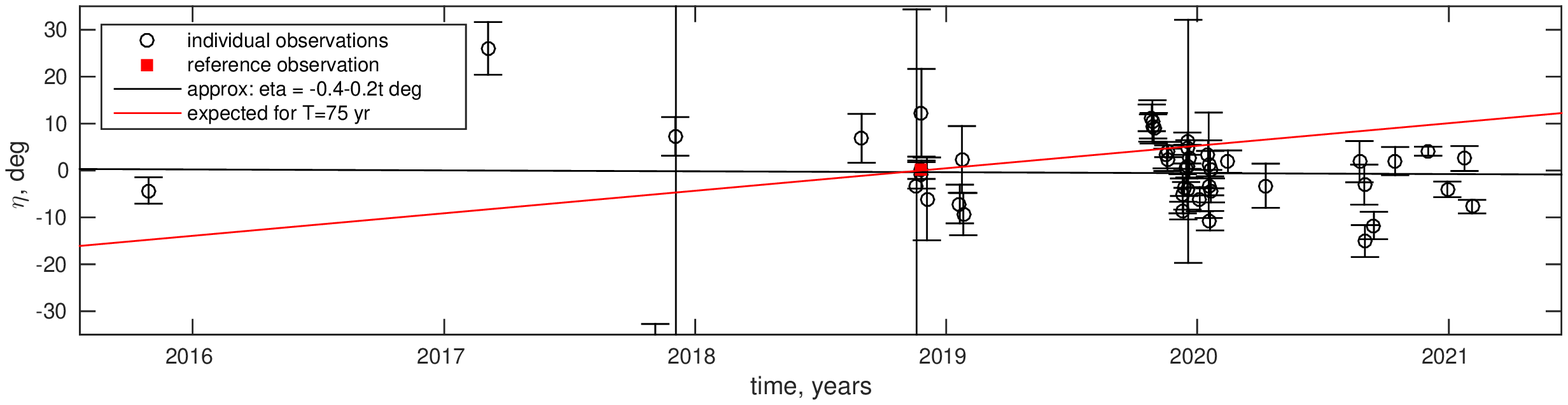}
\caption{The rotation angle $\eta$ determined from the approximation of individual observations with the reference observation (2018-11-25 23:49). Positive value means that the given observation is rotated CCW with respect to reference one. Black line indicates weighted linear fit: $\eta=\eta_0+\omega t$, where $\eta_0=-0.4\pm1.6^{\circ}$, $\omega=-0.2\pm1.1^{\circ}/$yr (68\% credible intervals). Red line indicates expected CCW rotation for $T=75$~yr.
\label{fig:rotation_time}}
\end{figure*}

Two spiral arms were detected in the disk of CQ~Tau earlier by \citet{Uyama2020} in $H$ band using Subaru/HiCIAO. The approximation of the spiral arms from their work is compared to our image in Fig.~\ref{fig:shadow}. As one can see, the brighter spiral found by \citet{Uyama2020} is slightly more distant from the star and at approximately the same position angle as in our images. Larger distance from the star is probably explained by the longer wavelength in case of HiCIAO observations: $1.6$~$\mu$m. Indeed, at the wavelength of our observation $0.8$~$\mu$m, the optical depth rises faster along the line of sight connecting the surface of the disk and the star. Thus one can expect that the region dominating in scattering light is closer to the star. The fainter arm from \citep{Uyama2020} at $0.3^{\prime\prime}$ from the star turned out to be inaccessible to us. DSP is less sensitive to fainter outer envelopes than adaptive optics assisted coronagraphs like HiCIAO due to less efficient separation of the radiation of bright central source and faint circumstellar material.

The comparison with ALMA data in 1.3~mm continuum by \citet{UbeiraGabellini2019} is presented in the same Fig.~\ref{fig:shadow}. In millimeter waves the envelope appears as a relatively large ring. The structures found by \citet{Uyama2020} and in present work are located inside this ring. Presumably the spiral in scattered light lies on the inner side of dust torus facing the star. We note that ALMA traces mm--sized dust particles, meanwhile visible and NIR radiation is scattered by much smaller particles. Spirals in scattered light inside a mm--wave dust ring  was reported for EM$^*$~SR~21 by \citep{Muro-Arena2020}. 

Palomar Testbed Interferometer observations of \citet{Eisner2004} in $K$ band demonstrated the existence of an inner disk around CQ~Tau with the characteristic size of 2.7~mas (gaussian fit), which corresponds to 0.45~AU at the distance of the source. The major axis of the inner disk has P.A.=$104\pm6^{\circ}$, the inclination of the disk is $48\pm5^{\circ}$. The disk with these parameters is indicated in Fig.~\ref{fig:shadow} by the red line. One can see that the orientations of the inner disk and the larger disk found by mm-wave observations are very different.

If the inner disk is geometrically thick, its inclination can be even larger. The fact that CQ~Tau is a UX Ori variable \citep{Berdyugin1989,Natta2000,Shakhovskoj2005} favours a larger inclination, the variable stars of this type display irregular brightness minimums induced by passages of dust clouds across the line of sight \citep{Grinin1988}. The minimums are accompanied by the rise of polarization. The UX Ori variability is typical for young stellar objects with highly inclined disks. On the basis of high resolution ESPaDOnS spectra \citet{Dodin2021} found recently that the clouds, occasionally eclipsing CQ Tau, has structures comparable in size with the stellar diameter.


\subsection{Is the spiral induced by a planet?}
\label{sec:temporal}

The pattern speed of a spiral induced by a planet is defined by the orbital motion of that planet. Therefore if the spiral detected by \citet{Uyama2020} and confirmed by our observations is indeed induced by the planet proposed by \citet{UbeiraGabellini2019}, its pattern speed should be $\approx4.8^{\circ}$/yr, which corresponds to orbital period of 75~yr. The direction of motion should be counter--clockwise.

To test this hypothesis we considered the following model of object variability. Firstly, as long as CQ~Tau is a UX~Ori variable, we allowed for a change in brightness and total polarization of an unresolved source. Secondly, we assumed that the envelope can rotate as a whole. For  rotation modeling we did not deproject the disk because, according to \citet{UbeiraGabellini2019}, its inclination is not so large: $35^{\circ}$ \citep{Dong2016}.

Using these assumptions we approximated each individual measurement of DPV with a reference one. For the reference observation we took the data obtained at 2018-11-25 23:49 which has relatively low noise level and lies roughly in the middle of considered time period. Note that this approximation employs the DPV measurements directly, not the images reconstructed using method from section~\ref{sec:polimg}. The approximation details are discussed in Appendix~\ref{app:approx}. In Appendix~\ref{app:approxdiff} we study the influence of choice of the reference observation on the result.

For each observation we obtained a set of parameters $q_\star^{\prime\prime}$, $u_\star^{\prime\prime}$, $\zeta$, $\eta$ by minimization of $\chi^2$ statistic. The first two of these parameters characterize the change in total polarization of the unresolved source, the third defines the change in its brightness. We interested most in parameter $\eta$, which is the rotation angle between the reference observation and a given one. The estimations of $\eta$ are presented in Fig.~\ref{fig:rotation_time}, as a function of time. 

In the same figure their weighted linear approximation by $\eta=\eta_0+\omega t$ is presented. $\eta$ values exhibit significant spread, which is probably due to variable shadowing of the envelope. However the overall trend is absent. The parameters of the linear fit are: $\eta_0=-0.4\pm1.6^{\circ}$, $\omega=-0.2\pm1.1^{\circ}$/yr (68\% credible intervals). Recall that the pattern speed expected for a planet--induced spiral is $\omega_e=4.8^{\circ}$/yr  (direction counter--clockwise), the corresponding trend is indicated by the red line in figure. Therefore we reject the hypothesis that the spiral in scattered light detected in our study is caused by a planet with the orbital period of 75~yr at the level of 4.3$\sigma$.

\subsection{The spiral as an effect of shadowing by the inner disk}
\label{sec:shadow}

The increased optical depth for the lines of sight passing near the equator of the inner disk should lead to the shadows on the outer disk. Such a mechanism was proposed to explain the appearance of disks in scattered light for a number of objects: HD100453 \citep{Benisty2017}, DoAr 44 \citep{Casassus2018}, RXJ1604.3-2130 \citep{Pinilla2018}. Given the small inclination of the outer disk, the shadow position should roughly coincide with the line of intersection of the inner and outer disks equator planes. 

\begin{figure}[ht]
\centering
\includegraphics[width=6.5cm]{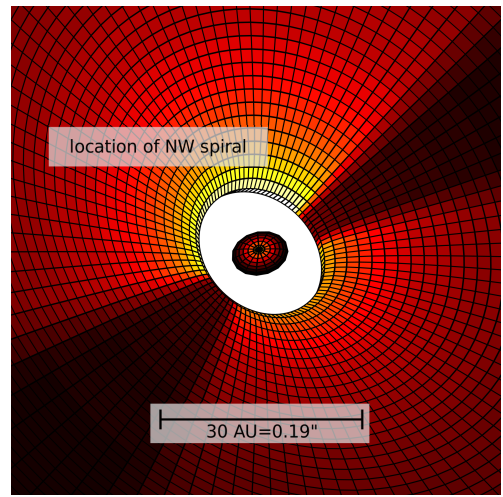}
\caption{Cartoon illustrating assumed mutual orientation of the outer and inner dust disks. The size of the inner disk is exaggerated. 
\label{fig:3Dshadow}}
\end{figure}

In our case, the brighter northern spiral arm indicates that the northern edge of the inner disk is farther from us. Taking into account the inclinations and position angles of the disks, the expected position angles of the shadow are $140^{\circ}$, $320^{\circ}$, as indicated in Fig.~\ref{fig:shadow}. Mutual configuration of the disks is illustrated in Fig.~\ref{fig:3Dshadow}. The angle between the axes of rotation of the inner and outer disks is $40^{\circ}$. The orientations of the shadows match the positions of the gaps in the spiral structure. Therefore the spirals detected by \citet{Uyama2020} and in the present work may be actually rings with shadows cast by the inner disk, rather than physical structures in scattering dust.


\section{Conclusions}
\label{sec:conclusions}

The structures found in the protoplanetary disks in scattered light show variability on timescales from days to years. Both physical motion \citep{Ren2018,Ren2020} and illumination changes \citep{Stolker2016,Benisty2017,Casassus2018,Pinilla2018} are observed. In each specific case it is important to differentiate these two basic explanations. The analysis of variations in the morphology of envelopes is greatly facilitated by multiepoch observations made by a homogeneous technique.

In the current paper, we present observations of the circumstellar envelope of a young star CQ~Tau by differential speckle polarimetry in the $I_\mathrm{c}$ band at the 2.5-m telescope of the Caucasian Mountain Observatory of Lomonosov Moscow State University, covering the period from 2015 to 2021. We detect a spiral arm at the characteristic distances $0.1-0.15^{\prime\prime}$ north of the star. The structure is located at the same position angle as the spiral arm discovered previously by \citet{Uyama2020} at a slightly larger stellocentric distance. We also report the presence of a southern, closer to the star, arm of the spiral.

To explain the large gap in the gaseous and dust disk of CQ Tau \citet{UbeiraGabellini2019} proposed the existence of a planet on the orbit with the semimajor axis of 20~AU. This possible planet could potentially generate structures appearing as spiral arms in scattered light, as was proposed for SAO 206462 \citep{Xie2021},  MWC 758 \citep{Ren2018}, AB Aur \citep{Boccaletti2020}. The pattern speed of the respective spiral in this case is dictated by the orbital motion of the planet: $4.8^{\circ}$/yr. On basis of observations covering the period of more than 5 yr we estimated the spiral pattern speed of CQ Tau to be $-0.2\pm1.1^{\circ}$/yr (68\% confidence interval). We conclude that the spiral in scattered light is not caused by that planet. However it is not excluded that the spiral is generated by a planet further away from the star, at distances of $\gtrsim$40-50~AU.

The relative stability of the spiral structure may indicate that its morphology is largely determined by shadowing of the outer disk by the misaligned inner one, like in cases of HD100453 \citep{Benisty2017}, DoAr 44 \citep{Casassus2018}, RXJ1604.3-2130 \citep{Pinilla2018}. The gaps in the spiral structure seen in scattered light at position angles of $\approx160^{\circ}$ and $\approx340^{\circ}$ correspond to the expected positions of the shadows from the inner disk found earlier by \citet{Eisner2004}. A slow precession of the inner disk is still possible. In the subsequent analysis and interpretation of the spiral structure of CQ Tau, it is important to consider the possibility of its non-uniform illumination.

\acknowledgments

We are grateful to Sergei Lamzin for the useful comments on manuscript. Constructive comments by anonymous referee allowed us to improve the analysis and presentation. We acknowledge the financial support of the Russian Science Foundation Public Monitoring Committee 20-72-10011. Scientific equipment used in this study was bought partially by the funds of the M. V. Lomonosov Moscow State University Program of Development.

%






\appendix

\section{Correction of instrumental polarization}
\label{app:instrpol}


The instrumental polarization induced by a mirror can be modelled using the formalism of Stokes vectors and Mueller matrices. The Stokes vector of outgoing radiation is the product of Stokes vector of incoming radiation and the Mueller matrix of the mirror. For the integral polarimetry the Mueller matrix of the mirror is averaged over the beam. Thus, mirrors acting in a symmetrical configuration, e.g. the primary and the secondary, do not modify the polarization of incoming radiation, except for a negligible depolarization. When the SPP is mounted in the Cassegrain focus the whole system up to beamsplitter possesses axial symmetry, thus ensuring low level of the instrumental polarization. Measurements of unpolarized stars demonstrated the instrumental polarization lower than $10^{-4}$ \citep{Safonov2017}.

In the Nasmyth focus an oblique reflection by the diagonal M3 mirror is added, axial symmetry is broken, and the instrumental polarization becomes quite substantial. It varies from 2 to 4\% over the wavelength range of SPP. In \citep{Safonov2017} we constructed a model of instrumental polarization induced by M3 mirror of 2.5-m telescope and provided a method for its correction.

\begin{deluxetable}{cc}
\tablenum{1}
\tablecaption{Instrumental polarization parameters in $I_c$ band, the Nasmyth focus of 2.5-m telescope of CMO SAI MSU. See the text for the definition of the symbols \label{tab:instrpol}}
\tablewidth{0pt}
\tablehead{\colhead{parameter} & \colhead{value}}
\decimalcolnumbers
\startdata
$q_\mathrm{ins}$, \% & $3.44\pm0.10$                \\
$u_\mathrm{ins}$, \% & $0\pm10$                  \\
$s_{q,\mathrm{ins}}^\star$, $\mu$as & $67\pm29$  \\
$t_{q,\mathrm{ins}}^\star$, $\mu$as & $702\pm65$ \\
$s_{u,\mathrm{ins}}^\star$, $\mu$as & $-344\pm19$\\
$t_{u,\mathrm{ins}}^\star$, $\mu$as & $37\pm18$  \\
\enddata
\end{deluxetable}

While the instrumental polarization is definitely important for measuring the integral polarization, for diffraction limited imaging it adds a new level of complexity. The fact that the reflection geometry is different for different points in the pupil lead to emergence of so called differential polarization aberrations \citep{Breckinridge2015}. The first order term of these aberrations manifests itself as a phase ramp in $\mathcal{R}_{Q,\mathrm{ins}}$ and $\mathcal{R}_{U,\mathrm{ins}}$, S19. The latter paper contains extensive analysis of instrumental terms $\mathcal{R}_{Q,\mathrm{ins}}$ and $\mathcal{R}_{U,\mathrm{ins}}$ and proposes the following model for them:
\begin{equation}
\mathcal{R}_{Q,\mathrm{ins}} = 1 + 2 q_\mathrm{ins} + i4\pi(s_{q,\mathrm{ins}}^\star f_x+t_{q,\mathrm{ins}}^\star f_y),
\label{eq:RinsQ}
\end{equation}
\begin{equation}
\mathcal{R}_{U,\mathrm{ins}} = 1 + 2 u_\mathrm{ins} + i4\pi(s_{u,\mathrm{ins}}^\star f_x+t_{u,\mathrm{ins}}^\star f_y).
\label{eq:RinsU}
\end{equation}
Here $f_x, f_y$ are the components of the two--dimensional frequency vector. $q_\mathrm{ins}$, $u_\mathrm{ins}$ is the polarization of an intrinsically unpolarized star detected by the instrument. $s_{q,\mathrm{ins}}^\star$, $t_{q,\mathrm{ins}}^\star$, $s_{u,\mathrm{ins}}^\star$, $t_{u,\mathrm{ins}}^\star$ are the components of so called instrumental polaroastrometric signal (IPS). Equations (\ref{eq:RinsQ},\ref{eq:RinsU}) assume that $\mathcal{R}_{Q,\mathrm{ins}}, \mathcal{R}_{U,\mathrm{ins}}$ are defined in the horizontal reference system.

\begin{figure*}[t!]
\centering
\includegraphics[width=15cm]{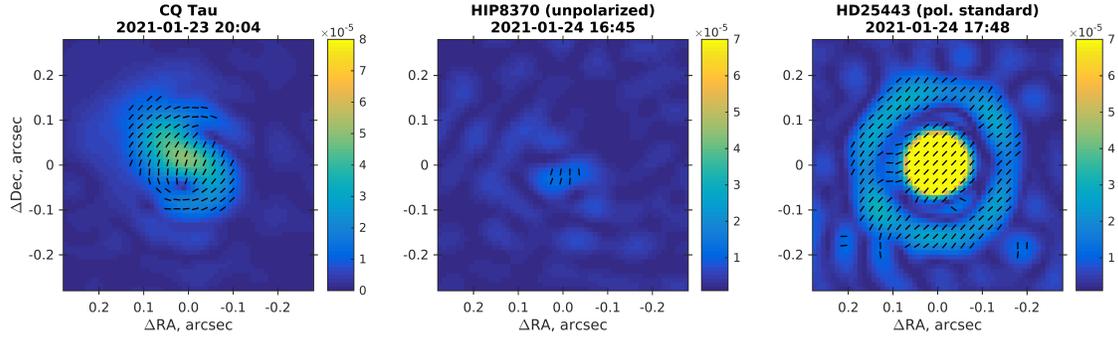}
\caption{The same as in 5th column of Fig.~\ref{fig:rvals}, but for observations obtained in the Nasmyth focus. From left to right, CQ Tau, an unpolarized star, a polarization standard. The date of respective observation is indicated in the title.
\label{fig:polimgNasmyth}}
\end{figure*}

The quantities $q_\mathrm{ins}$, $u_\mathrm{ins}$, $s_{q,\mathrm{ins}}^\star$, $t_{q,\mathrm{ins}}^\star$, $s_{u,\mathrm{ins}}^\star$, $t_{u,\mathrm{ins}}^\star$ were determined in S19 from measurements of point--like unpolarized stars, see their values for $I_c$ band in Table \ref{tab:instrpol}. In the Nasmyth focus the IPS is large and should be corrected through the division of observed $\mathcal{R}_{c4}$ and $\mathcal{R}_{s4}$ by (\ref{eq:RinsQ}) and (\ref{eq:RinsU}), respectively. This operation is performed in the horizontal reference system.

Fig.~\ref{fig:polimgNasmyth} demonstrates the same quantities as in the fifth column of Fig.~\ref{fig:rvals}, but obtained in the Nasmyth focus and corrected for the instrumental polarization. Again for comparison we give the observations for a point--like unpolarized star and a point--like polarization standard, HIP8370 and HD25443, respectively. One can see that the spiral feature is persistent for the CQ Tau. It is not observed neither for the unpolarized star, nor for the polarization standard. 

The image of the polarization standard HD25443 exhibits a ring in polarized light, which corresponds to the first bright ring of the Airy function. Note that this ring has a constant polarization angle equal to the integral polarization angle of the star. In this respect it differs from the azimuthal polarization pattern, as in the case of CQ Tau, which is a genuine feature of a scattering reflection nebula.

\section{Approximation of one observation by another}
\label{app:approx}

\begin{figure}[b]
\centering
\begin{minipage}[b]{0.45\linewidth}
\includegraphics[width=9cm]{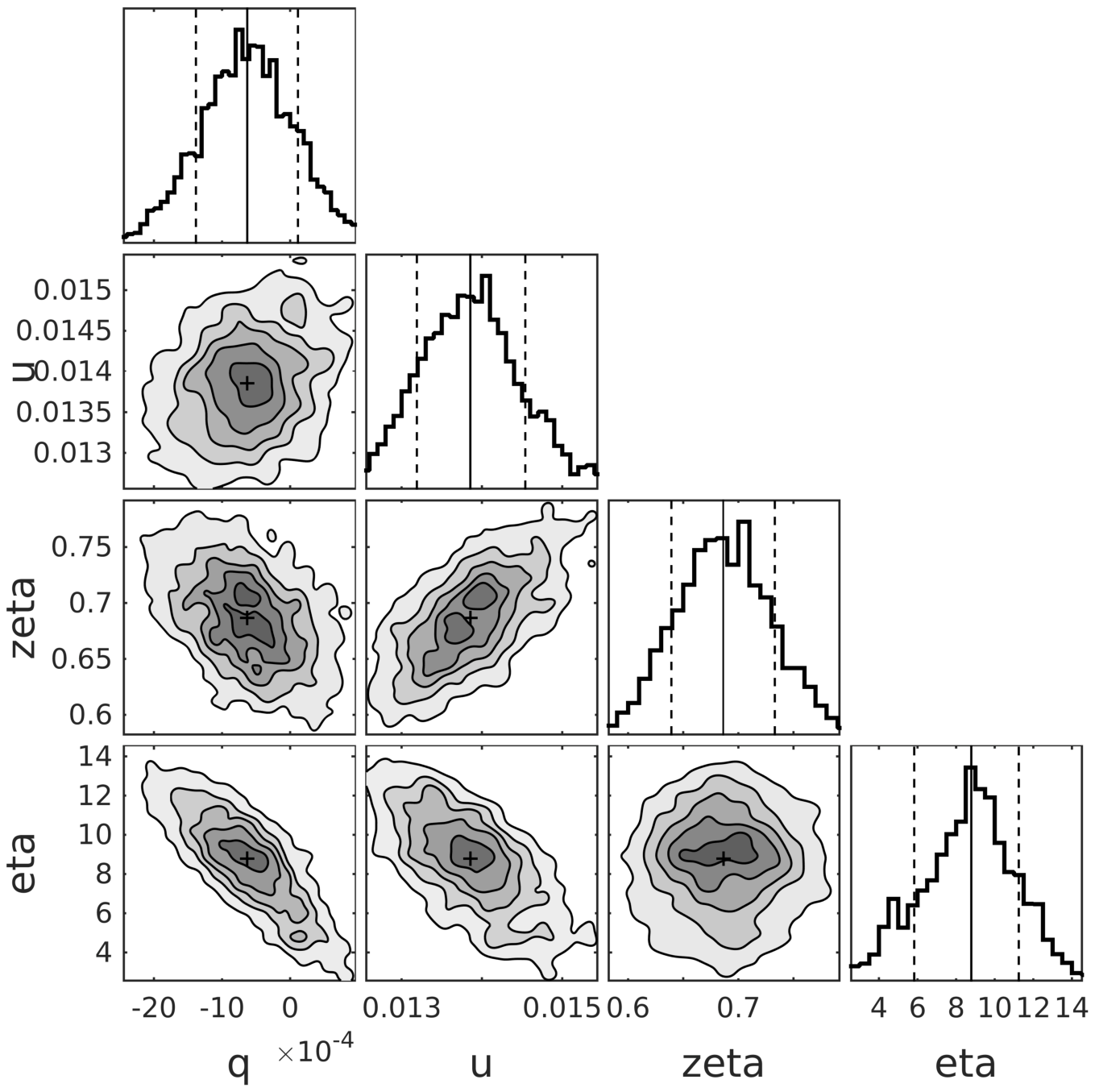}
\end{minipage}\quad
\begin{minipage}[b]{0.45\linewidth}
\centering
\includegraphics[width=9cm]{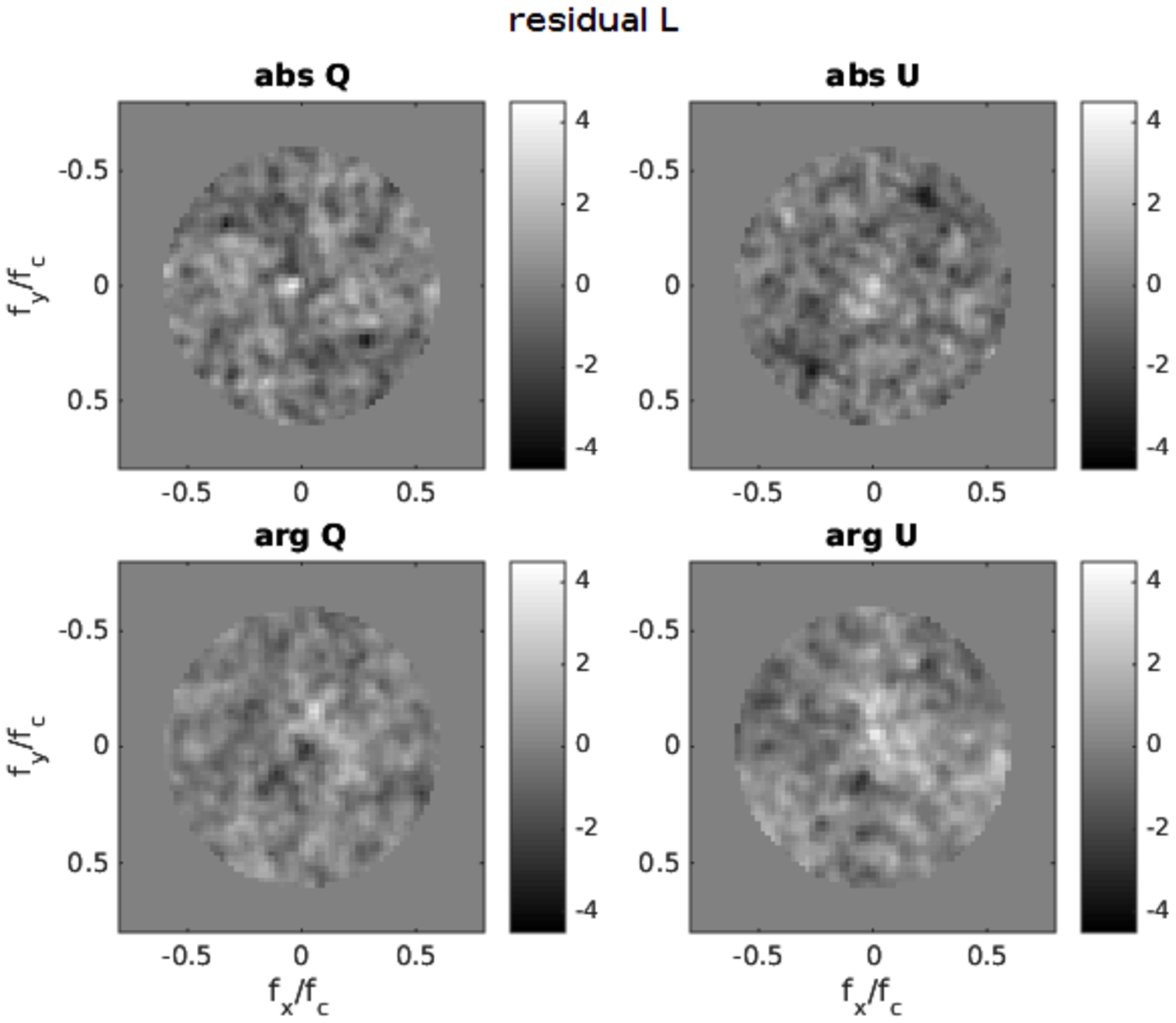}
\end{minipage}
\caption{The approximation of observation obtained at 2019-10-26 22:10 by the reference one (2018-11-25 23:49). {\it Left panel:} Corner plot of the marginalized posterior distributions for the model parameters: $q_\star^{\prime\prime}$, $u_\star^{\prime\prime}$, $\zeta$, $\eta$. The posterior was sampled by the Markov chain Monte Carlo method, see the text for details. Contours in the two-dimensional distributions are for the ratio of probability to its maximum value and correspond to 0.9, 0.7, 0.5, 0.3, 0.1. The optimal values of parameters and 68\% confidence intervals are indicated in one-dimensional distributions by the solid and dashed lines, respectively. {\it Right panel:} The components of the residuals defined by equations (\ref{eq:residuals}), for the best fit indicated in the left panel. Absolute and phase parts of $\mathcal{L}_Q$ and $\mathcal{L}_U$ are given in upper and lower rows, respectively.
\label{fig:cornerplot}}
\end{figure}

The section \ref{sec:temporal} is based on the approximation of a given observation $k$ by a reference one $j$. The differential polarization visibility for the reference observation can be written as:
\begin{equation}
\mathcal{R}_Q^j(\boldsymbol{f}) = 1 + 2q_\star + 2I_\star^{-1}\widetilde{Q}_\mathrm{e}(\boldsymbol{f}),
\label{eq:app_DPV_Q1}
\end{equation}
according to equations (\ref{eq:objectI},\ref{eq:objectQ}). Recall that $q_\star$ is the polarization of the star, $Q_\mathrm{e}$ is the Fourier transform of envelope image in Stokes parameter $Q$, $I_\star$ is the stellar flux. If the envelope rotated by $\eta$ (positive for counter--clockwise rotation), then the DPV will be:
\begin{equation}
\mathcal{R}_{Q,r}^{j}(\eta;\boldsymbol{f}) = 1 + 2q_\star + 2I_\star^{-1}\widetilde{Q}_{\mathrm{e},r}(\eta;\boldsymbol{f}).
\label{eq:app_DPV_Q2}
\end{equation}
The rotated version of $\widetilde{Q}_{\mathrm{e},r}$ will depend on both $\widetilde{Q}_{\mathrm{e}}$ and $\widetilde{U}_{\mathrm{e}}$ due to transformation of Stokes parameters in the rotated reference system. For the equations describing rotation of $\mathcal{R}_Q$ and $\mathcal{R}_U$ see appendix B of S19.

Suppose that the stellar flux increases by $\zeta$ and stellar polarization changes to $q_\star^{\prime}$. The DPV will change in the following way:
\begin{equation}
\mathcal{R}_{Q,r}^{j*}(\eta;\boldsymbol{f}) = 1 + 2q_\star^{\prime} + 2(\zeta I_\star)^{-1}\widetilde{Q}_{\mathrm{e},r}(\eta;\boldsymbol{f}).
\label{eq:app_DPV_Q3}
\end{equation}

Expressing $\widetilde{Q}_\mathrm{e}$ from (\ref{eq:app_DPV_Q2}) and substituting it in (\ref{eq:app_DPV_Q3}), we obtain formula describing the transformation of the   measurement:
\begin{equation}
\mathcal{R}_{Q,r}^{j*}(\eta;\boldsymbol{f}) = 1+2\bigr[q_\star^{\prime}-\zeta^{-1}q_\star \bigr]+\zeta^{-1}\bigl[\mathcal{R}^{j}_{Q,r}(\eta;\boldsymbol{f})-1\bigr],
\label{eq:RmodQ}
\end{equation}
Similar relation can be formulated for Stokes $U$: 
\begin{equation}
\mathcal{R}_{U,r}^{j*}(\eta;\boldsymbol{f}) = 1+2\bigr[u_\star^{\prime}-\zeta^{-1}u_\star \bigr]+\zeta^{-1}\bigl[\mathcal{R}^{j}_{U,r}(\eta;\boldsymbol{f})-1\bigr],
\label{eq:RmodU}
\end{equation}
The second terms in these equations will be considered as new parameters: $q_\star^{\prime\prime}=q_\star^{\prime}-\zeta^{-1}q_\star$,  $u_\star^{\prime\prime}=u_\star^{\prime}-\zeta^{-1}u_\star$.

The comparison of measurements is performed by computation of the residual of the following form:
\begin{equation}
\mathcal{M}(\boldsymbol{f}_i)=\mathcal{L}_{Q}^2(\boldsymbol{f}_i)+\mathcal{L}_{U}^2(\boldsymbol{f}_i),
\label{eq:residual}
\end{equation}
where
\begin{equation}
\mathcal{L}_{Q}(\boldsymbol{f}_i)\,=\,  \frac{\mathcal{R}_{Q}^k(\boldsymbol{f}_i)-\mathcal{R}_{Qr}^{j*}(\boldsymbol{f}_i)}{\sqrt{\sigma_{R,k}^2(\boldsymbol{f}_i)+\sigma_{R,j*}^{2}(\boldsymbol{f}_i)}},\,\,\,\,\mathcal{L}_{U}(\boldsymbol{f}_i)\,=\,  \frac{\mathcal{R}_{U}^k(\boldsymbol{f}_i)-\mathcal{R}_{Ur}^{j*}(\boldsymbol{f}_i)}{\sqrt{\sigma_{R,k}^2(\boldsymbol{f}_i)+\sigma_{R,j*}^{2}(\boldsymbol{f}_i)}}
\label{eq:residuals}
\end{equation}
Here $\mathcal{R}_{Q}^k$, $\mathcal{R}_{U}^k$ are the DPV measurements for the observation $k$, $\mathcal{R}_{Q,r}^{j*}$, $\mathcal{R}_{U,r}^{j*}$ are the same, but for observation $j$, modified according to equations (\ref{eq:RmodQ}) and (\ref{eq:RmodU}). $\sigma_{R,k}, \sigma_{R,j*}$ are the errors of estimation of DPV, the method of their determination can be found in Appendix B of S19.

For each observation we determine four parameters: $q_\star^{\prime\prime}$, $u_\star^{\prime\prime}$, $\zeta$, $\eta$ by minimization of the total residual:
\begin{equation}
\sum_{|f_i|<0.7f_c} \mathcal{M}(\boldsymbol{f}_i).
\label{eq:totresid}
\end{equation}
Here the summation takes place for the frequencies smaller than $0.7f_c$, where $f_c$ is the cut--off frequency.  Note that the total residual defined by (\ref{eq:totresid}) is technically a $\chi^2$ statistic. The posterior probability of parameters was sampled using the affine-invariant ensemble Markov Chain Monte Carlo method \citep{Goodman2010}. 50 walkers and $2\times10^4$ iterations in total were used. 

\begin{figure}[b]
\centering
\includegraphics[width=15cm]{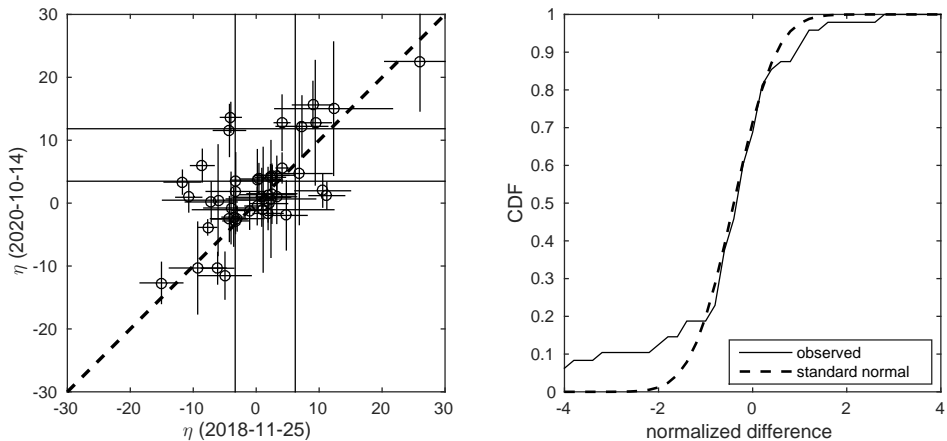}
\caption{{\it Left panel:} comparison of rotation angles $\eta$ for individual observations with respect to two reference observation. Along horizontal axis $\eta_1$ evaluated relative to observation at 2021-11-25, along vertical axis $\eta_2$ evaluated relative to observation 2020-10-14 (see section \ref{sec:temporal} and appendix \ref{app:approx} for details of approximation). {\it Right panel:} Solid line --- cumulative distribution function of differences $\eta_1-\eta_2$, normalized by the uncertainties. Dashed line --- normal distribution with variance equal to unity and mean equal to -0.4, see the text for the discussion.
\label{fig:rotation_compar}}
\end{figure}

The example of the resulting posterior for observation 2019-10-26 22:10 approximated by 2018-11-25 23:49 is displayed in the left part Fig.~\ref{fig:cornerplot}.  The best values for parameters are: $q_\star^{\prime\prime}=-0.06\pm0.07$\%, $u_\star^{\prime\prime}=-1.32\pm0.07$\%, $\zeta=0.67\pm0.05$, $\eta=8.8\pm2.7$ (values after the plus--minus sign are 68\% confidence intervals). The components of the corresponding residual (\ref{eq:residuals}) are presented in the right part of Fig. \ref{fig:cornerplot}. The absence of significant departures indicates that the fitting of one observation by modified version of another has satisfactory quality.

\section{The dependence of results on reference observation selection}
\label{app:approxdiff}

In section \ref{sec:temporal} all observations were compared to a reference one in order to search for the rotation of the envelope. Here we consider how the choice of the reference observation may affect the results of this analysis. For an alternative reference observation we took the data obtained on 2020-10-14 23:00. Then the procedure described in section \ref{sec:temporal} was repeated for this reference.

The resulting rotation angles $\eta_1$ and $\eta_2$ for two references are directly compared in the left panel of the Fig.~\ref{fig:rotation_compar}. The CDF of $\eta_1-\eta_2$ normalized by its uncertainty is presented in the right panel of the same figure. These normalized deviations are expected to follow shifted standard normal distribution. The shift emerges because the envelope may have rotated between moments 2018-11-25 23:49 and 2020-10-14 23:00. One can see that the distribution of deviations indeed follows the normal distribution for the majority of data points. However the wings of the distribution are larger than expected.

The dependence of rotation angles $\eta$, determined using the alternative reference observation, on time were fitted with the linear law, as it was done in section~\ref{sec:temporal}. The parameters of the fit are $\eta_0=1.4\pm1.7^{\circ},\omega=-0.7\pm0.9^{\circ}$/yr (68\% credible intervals). They coincide within errors with the values obtained using the observation for 2018-11-25 23:49 as a reference. We conclude that the choice of the reference observation has no major effect on the results of section \ref{sec:temporal}.

\section{Observations of CQ Tau}

\startlongtable
\begin{deluxetable*}{cccccccccc}
\tablenum{2}
\tablecaption{Observational log\label{tab:obslog}}
\tablewidth{0pt}
\tablehead{
\colhead{focal} & \colhead{observation} & \colhead{number of} & \colhead{$p$} & \colhead{$\chi$} & \colhead{$I_\mathrm{c}$} & \colhead{$\sigma$} & \colhead{$z$} & \colhead{$h_{\odot}$} & \colhead{$\beta$}\\
\colhead{station}      & \colhead{UT}   & \colhead{frames}       & \colhead{(\%)} & \colhead{($^{\circ}$)} & \colhead{(mag)} & \colhead{} & \colhead{($^{\circ}$)} & \colhead{($^{\circ}$)} & \colhead{($^{\prime\prime}$)}
}
\decimalcolnumbers
\startdata
C & 2015-10-29 00:51 & 40029 &$0.82\pm0.01$ & $78.4\pm0.2$ &  --- &  0.0065&  20 & -31 & --- \\ 
C & 2017-03-05 18:29 & 9131 &$0.85\pm0.01$ & $164.6\pm0.4$ &  --- &  0.0227&  37 & -37 & 1.31 \\ 
N & 2017-11-03 22:06 & 10237 &$0.30\pm0.15$ & $147.8\pm14.4$ &  --- &  0.0109&  29 & -58 & 1.07 \\ 
C & 2017-12-02 22:55 & 10131 &$1.12\pm0.03$ & $177.2\pm0.8$ &  --- &  0.0091&  22 & -58 & 0.92 \\ 
N & 2017-12-03 20:42 & 9960 &$0.90\pm0.15$ & $168.7\pm4.8$ &  --- &  0.0106&  24 & -68 & 0.87 \\ 
N & 2018-08-31 01:23 & 27535 &$0.65\pm0.15$ & $163.2\pm6.6$ &  --- &  0.0106&  39 & -12 & 1.01 \\ 
N & 2018-11-18 23:50 & 5081 &$0.86\pm0.15$ & $9.3\pm5.0$ &  --- &  0.0976&  22 & -46 & 1.16 \\ 
N & 2018-11-25 23:33 & 29939 &$0.92\pm0.15$ & $152.0\pm4.7$ &  --- &  0.0075&  23 & -50 & 0.88 \\ 
N & 2018-11-25 23:49 & 29916 &$0.89\pm0.15$ & $152.2\pm4.8$ & $9.3\pm0.1$ &  0.0071&  25 & -47 & --- \\ 
N & 2018-11-26 00:11 & 2937 &$0.83\pm0.15$ & $152.8\pm5.2$ &  --- &  0.0338&  29 & -44 & --- \\ 
N & 2018-12-03 22:57 & 6437 &$0.70\pm0.15$ & $166.7\pm6.1$ &  --- &  0.0145&  23 & -57 & 1.06 \\ 
C & 2019-01-20 19:17 & 12031 &$1.35\pm0.03$ & $147.6\pm0.6$ &  $9.7\pm0.1$ &  0.0126&  20 & -55 & 1.00 \\ 
N & 2019-01-23 21:28 & 11944 &$0.82\pm0.15$ & $151.8\pm5.3$ &  --- &  0.0333&  40 & -66 & 1.22 \\ 
N & 2019-01-25 19:54 & 11934 &$0.41\pm0.15$ & $158.7\pm10.6$ &  --- &  0.0152&  26 & -59 & 0.98 \\ 
N & 2019-10-26 22:10 & 11938 &$0.87\pm0.15$ & $29.0\pm4.9$ &  $8.7\pm0.1$ &  0.0087&  34 & -55 & 0.88 \\ 
N & 2019-10-27 22:09 & 11960 &$0.57\pm0.15$ & $49.5\pm7.6$ &  $8.5\pm0.1$ &  0.0085&  33 & -55 & 0.57 \\ 
N & 2019-10-28 21:46 & 11939 &$0.70\pm0.15$ & $27.4\pm6.1$ &  $8.9\pm0.1$ &  0.0077&  36 & -57 & 0.53 \\ 
N & 2019-10-29 20:23 & 11939 &$1.44\pm0.15$ & $11.4\pm3.0$ &  $9.2\pm0.1$ &  0.0145&  50 & -59 & 0.81 \\ 
N & 2019-11-17 21:06 & 11950 &$1.02\pm0.15$ & $54.2\pm4.2$ &  $9.6\pm0.1$ &  0.0291&  30 & -65 & 1.62 \\ 
N & 2019-11-18 21:12 & 11934 &$0.56\pm0.15$ & $83.0\pm7.7$ &  $9.8\pm0.1$ &  0.0144&  28 & -65 & 1.25 \\ 
N & 2019-11-20 01:29 & 11948 &$1.26\pm0.15$ & $135.8\pm3.4$ &  $9.8\pm0.1$ &  0.0230&  37 & -29 & 0.83 \\ 
N & 2019-12-10 21:36 & 11941 &$0.91\pm0.15$ & $110.7\pm4.7$ &  $9.0\pm0.1$ &  0.0064&  19 & -68 & 0.89 \\ 
C & 2019-12-11 22:16 & 12047 &$1.56\pm0.04$ & $156.9\pm0.8$ &  $9.8\pm0.1$ &  0.0101&  22 & -64 & 0.74 \\ 
C & 2019-12-12 23:28 & 12948 &$1.21\pm0.04$ & $159.9\pm1.0$ &  $9.6\pm0.1$ &  0.0137&  32 & -54 & 0.90 \\ 
N & 2019-12-15 21:44 & 12943 &$1.09\pm0.15$ & $111.9\pm3.9$ &  $9.3\pm0.1$ &  0.0178&  20 & -68 & 1.44 \\ 
N & 2019-12-17 18:41 & 12958 &$0.79\pm0.15$ & $95.5\pm5.4$ &  $8.7\pm0.1$ &  0.0086&  34 & -54 & 0.74 \\ 
N & 2019-12-17 20:27 & 12950 &$1.10\pm0.15$ & $99.6\pm3.9$ &  $8.6\pm0.1$ &  0.0090&  20 & -68 & 1.07 \\ 
N & 2019-12-17 23:42 & 8384 &$1.06\pm0.15$ & $114.6\pm4.1$ &  $8.6\pm0.1$ &  0.0293&  38 & -52 & --- \\ 
N & 2019-12-18 23:07 & 1933 &$1.01\pm0.15$ & $116.6\pm4.3$ &  $8.4\pm0.0$ &  0.7559&  33 & -58 & 2.34 \\ 
N & 2019-12-20 18:53 & 12945 &$0.21\pm0.15$ & $0.6\pm20.6$ &  $9.2\pm0.1$ &  0.0324&  30 & -56 & 1.58 \\ 
N & 2020-01-04 19:41 & 12939 &$2.13\pm0.15$ & $93.4\pm2.0$ &  $9.4\pm0.1$ &  0.0062&  19 & -62 & 1.03 \\ 
N & 2020-01-16 22:13 & 12937 &$1.31\pm0.15$ & $6.2\pm3.3$ &  $9.9\pm0.1$ &  0.0501&  43 & -64 & 1.25 \\ 
N & 2020-01-17 18:43 & 12950 &$2.10\pm0.15$ & $22.2\pm2.0$ &  $10.9\pm0.1$ &  0.5447&  19 & -50 & 1.44 \\ 
N & 2020-01-19 15:49 & 12948 &$2.38\pm0.15$ & $33.1\pm1.8$ &  $9.8\pm0.1$ &  0.0205&  41 & -19 & 1.11 \\ 
N & 2020-01-19 18:17 & 12938 &$2.05\pm0.15$ & $37.7\pm2.1$ &  $10.0\pm0.1$ &  0.0093&  20 & -46 & 0.90 \\ 
N & 2020-01-19 21:44 & 12954 &$1.73\pm0.15$ & $24.9\pm2.5$ &  $10.0\pm0.1$ &  0.0314&  40 & -66 & 0.92 \\ 
N & 2020-01-20 22:19 & 12962 &$0.98\pm0.15$ & $72.5\pm4.4$ &  $9.6\pm0.1$ &  0.0094&  47 & -63 & 0.94 \\ 
N & 2020-02-14 19:26 & 12940 &$2.34\pm0.15$ & $101.2\pm1.8$ &  $9.0\pm0.1$ &  0.0106&  34 & -50 & 0.83 \\ 
N & 2020-04-09 17:15 & 12939 &$1.35\pm0.15$ & $92.3\pm3.2$ &  $8.8\pm0.1$ &  0.0173&  49 & -16 & 0.70 \\ 
N & 2020-08-24 01:11 & 12955 &$0.37\pm0.15$ & $12.4\pm11.6$ &  --- &  0.0163&  45 & -13 & 0.90 \\ 
N & 2020-08-31 01:22 & 19549 &$0.31\pm0.15$ & $63.0\pm13.6$ &  $8.6\pm0.1$ &  0.0089&  39 & -13 & 1.15 \\ 
N & 2020-09-01 01:14 & 59933 &$0.37\pm0.15$ & $57.0\pm11.6$ &  $8.7\pm0.1$ &  0.0053&  39 & -14 & 0.86 \\ 
N & 2020-09-14 02:07 & 54937 &$0.39\pm0.15$ & $16.0\pm11.1$ &  $9.0\pm0.1$ &  0.0032&  23 & -8 & 0.87 \\ 
N & 2020-10-14 23:00 & 34948 &$0.97\pm0.15$ & $11.1\pm4.4$ &  $8.7\pm0.1$ &  0.0039&  32 & -45 & 0.57 \\ 
N & 2020-12-02 00:23 & 34944 &$1.17\pm0.15$ & $146.3\pm3.7$ &  $9.6\pm0.1$ &  0.0088&  35 & -43 & 0.95 \\ 
N & 2020-12-29 19:58 & 35346 &$0.92\pm0.15$ & $114.9\pm4.7$ &  $8.9\pm0.1$ &  0.0062&  19 & -65 & 0.66 \\ 
N & 2021-01-23 20:04 & 34953 &$0.59\pm0.15$ & $120.9\pm7.3$ &  $8.9\pm0.1$ &  0.0044&  27 & -61 & 0.80 \\ 
N & 2021-02-03 17:40 & 34927 &$0.80\pm0.15$ & $16.9\pm5.4$ &  $9.3\pm0.1$ &  0.0070&  19 & -36 & 0.76 \\ 
\enddata
\tablecomments{Focal stations: C --- Cassegrain, N --- Nasmyth. $p$ and $\chi$ are the fraction and angle of polarization, respectively. $I_c$ band magnitude estimated by quasi--simultaneous observations of HIP25001. At some epochs these observation were not conducted. $\sigma$ is uncertainty of $\mathcal{R}$ averaged over region $0.3f_c<|\boldsymbol{f}|<0.4f_c$, see Section~\ref{sec:observations}. $z$ is zenith angle, $h_\odot$ is the Sun altitude, $\beta$ is seeing measured by MASS--DIMM instrument \citep{Kornilov2014}. The exposure of a single frame was 30~ms for all observations. All observations were conducted in $I_\mathrm{c}$ band, which effective wavelength is 0.806~nm.}
\end{deluxetable*}


\bibliography{references}{}
\bibliographystyle{aasjournal}



\end{document}